\documentstyle[psfig,conf_iap,10pt]{article}

\def\spose#1{\hbox to 0pt{#1\hss}}
\def\simlt{\mathrel{\spose{\lower 3pt\hbox{$\mathchar"218$}}
     \raise 2.0pt\hbox{$\mathchar"13C$}}}
\def\simgt{\mathrel{\spose{\lower 3pt\hbox{$\mathchar"218$}}
     \raise 2.0pt\hbox{$\mathchar"13E$}}}
\def\lsim{\rlap{$<$}{\lower 1.0ex\hbox{$\sim$}}}
\def\gsim{\rlap{$>$}{\lower 1.0ex\hbox{$\sim$}}}

\def\kms{\mbox{{\rm km~s}$^{-1}$}}

\newcommand{\Lya}{\mbox{Ly$\alpha$}}
\newcommand{\Lyb}{\mbox{Ly$\beta$}}

\newcommand{\civ}{\mbox{C {\sc iv}}}

\begin{document}
\heading{
%
%
10 Mpc QSO Absorber Correlations at $z\sim 3$
%
}
\par\medskip\noindent
\author{%
G.M. Williger$^1$, A. Smette$^1$, C. Hazard$^{2,3}$, J.A. Baldwin$^4$, R.G.
McMahon$^2$
}
\address{%
Code 681, NASA/GSFC, Greenbelt MD 20771 USA
}
\address{%
Institute of Astronomy, Madingley Rd, Cambridge CB3 0HA, England
}
\address{%
Dept. of Physics \& Astronomy, Univ. Pittsburgh, Pittsburgh PA 15260 USA
}
\address{%
CTIO, Casilla 603, La Serena, Chile
}

\begin{abstract}
  We present results from a survey of the \Lya\ forest at
  $2.15<z<3.26$ toward ten QSOs concentrated within a 1$^\circ$ field.
  We find correlations of the \Lya\ absorption line wavelengths
  between different lines-of-sight over the whole redshift range.
  This indicates the existence of large-scale structures in the
  Ly-$\alpha$ forest extending at least over $\sim 36 ~ h^{-1}$
  comoving Mpc in the plane of the sky, as may be expected from recent
  large scale structure simulations.
\end{abstract}
\section{Introduction}

Comparison between spectra of multiply lensed quasars
{\cite{Smette95,Smette92}} or close quasar pairs
{\cite{Bechtold94b,Dinshaw95,Dinshaw94,Fang96}} indicate that the
numerous narrow \Lya\ absorption lines observed in quasar spectra are
produced in large tenuous clouds with diameters $>50$ $h^{-1}$ kpc
($H_0\equiv 100 h$ \kms\ Mpc$^{-1}, q_0=0.5$ and $\Lambda=0$ assumed
throughout).  Their association with galaxies is unclear
{\cite{Bowen96,Lanzetta95,LeBrun96,Rauch96}}.  There has been much
effort made to examine spatial structure in the \Lya\ forest along
isolated lines-of-sight ({\cite{Kirkman97}} and references therein).  A
complementary approach is to examine structure between adjacent
lines-of-sight, which has already revealed structures on the scale of
several Mpc outlined by \civ\ absorbers {\cite{Dinshaw96,Williger96}}.
Crotts {\cite{Crotts89}} searched for spatial structure in the \Lya\
forest at $2.2<z<2.6$ on the scales of several arcmin separation using
4 QSOs separated by at most 440 arcsec.  He found clustering for
velocity separations of $\Delta v \approx 100$ \kms\ for $W_0 \geq 0.4$
\AA\ absorbers and for $\Delta v \approx 400$ \kms\ for stronger
absorbers ($W_0 \geq 0.8$ \AA). A new analysis method, using pixel
correlations and based on the same SGP data as used here, reveals the
\civ\ cluster at $z\sim 2.3$ which was found in {\cite{Williger96}}, as
well as a void toward four lines-of-sight at least $36\times 24 h^{-2}$
comoving Mpc in extent at $z=2.97$ {\cite{Liske97}}.

Here we present the 2-point correlation function of the \Lya\ forest
toward ten $z>2.3$ QSOs within a $1^\circ$ field near the south
Galactic pole. A more complete description of the statistical methods
and results of this work will be presented elsewhere
{\cite{Williger97}}.

\section{Data}

The observational data consist of the 10 highest signal-to-noise ratio
2 \AA\ resolution spectra covering the Ly-$\alpha$ forest; they were
obtained during a parallel study {\cite{Williger96}} on the large scale
structure revealed by C\,{\sc iv} absorbers.

We exclude lines within $5000$ \kms\ from the background QSO to avoid
uncertainties linked with the ``proximity effect'' (reduced line
number density and equivalent widths), and all \Lya\ lines
corresponding to known metal absorption systems from
{\cite{Williger96}}.  Apart of these restrictions, we use all of the
\Lya\ absorber information we have between \Lya\ and \Lyb\ emission
wavelengths, regardless of the signal-to-noise ratio of the data.

The sample contains 383 \Lya\ lines at $2.15<z<3.26$, at a rest
equivalent width detection limit of $W_0 \geq 0.1$ \AA\ (5$\sigma$).
The data are only complete to $W_0 = 0.5$ \AA.  The number density of
$W_0\geq 0.5$ \AA\ lines is typical, being proportional to a power
law{\cite{Lu91}} $d{\cal N}/dz\propto (1+z)^\gamma$, $\gamma=2.5\pm
0.8$. We find no large voids at any $W_0$ (following
{\cite{Ostriker88}}).

\section{ Statistical analysis}

We first calculate the 2-point correlation function of the comoving
separation of the \Lya\ absorbers.  The comoving separations are
evaluated from the redshift of the absorption lines and the separation
between the lines-of-sight using the standard formula\cite{Crotts89}.
In order to estimate the significance of any signal, we constructed
control samples free of correlations between absorbers along different
lines-of-sight by shifting the observed absorption line wavelengths
along each line-of-sight by a random amount varying between 6 and
182~\AA (or 2 and 52 $h^{-1}$ comoving Mpc or 420 and 11000 \kms\
along the line of sight at $z\approx 3.25$).  This procedure allows us
to use a much larger line sample than from a complete survey, which
would be limited to only 149 lines at $W_0>0.5$ \AA\ for $5\sigma$
significance, while keeping the observational limitations of spectral
resolution and signal-to-noise ratio.
The dispersion about the mean per bin was measured
directly from the Monte Carlo simulations, typically using 100--1000
iterations.  Monte Carlo simulations show that the distribution of
pair counts in a given bin is well-approximated by a Gaussian.  We
find no evidence (for any binning) for three dimensional structure in
the \Lya\ forest using the two point correlation function over $2\sim
52 h^{-1}$ Mpc.

We next calculate the 2-point correlation function of the velocity
differences of all possible pair of \Lya\ lines, testing for
structures which are in the plane of the sky.  We only include
absorber pairs which lie along {\it different} lines-of-sight, in
order to avoid resonances between \Lya\ and unidentified metal lines
in the same spectrum, and to concentrate on structures with transverse
spatial extent.  We find an excess of pairs of $W_0\geq 0.1$ \AA\
\Lya\ absorption lines with velocity differences $50 < \Delta v / {\rm
  km\, s}^{-1} < 100$ compared to the control sample at the
3.3$\sigma$ level.  It is noteworthy that we do not detect any
significant signal at velocity splittings $\Delta v<50$ \kms\ under
any circumstances.  The signal is not present at $2.15<z<2.60$, and is
strongest at $2.60<z<3.25$ for lines $0.1<W_0/{\rm \AA\ }<0.7$ (Fig.
1), at the 3.7$\sigma$ level, with the probability of finding such an
excess in {\it any} bin of $P=0.001$. We find no significant dependance
of the correlation function on the angular separation.

\begin{figure}
\centerline{\vbox{
\psfig{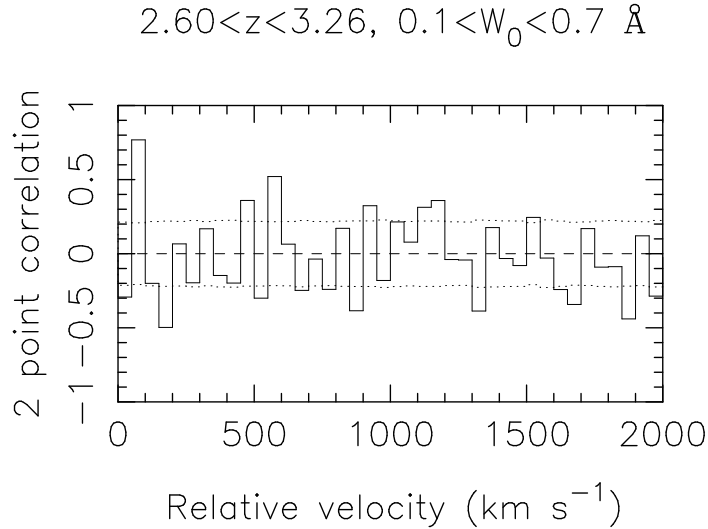}
}}
\caption[]{The two-point correlation function in velocity
space for the
SGP data at $2.60<z<3.25$ and $0.1<W_0<0.7$ \AA .  Only absorber pairs
along {\it different} lines-of-sight are
included.
The $1\sigma$ scatter measured from fluctuations in the randomized data is
shown by the dotted lines.  The $50<\Delta v<100$ \kms\ feature
is significant
at $3.7\sigma$, with a probability of occurring in {\it any} bin
of $P=0.001$.

}
\end{figure}

%
\section{Discussion }

The most intensive previous study for correlation between close QSO sightlines
used 4
QSOs{\cite{Crotts89}}, with similar resolution and slightly lower
signal-to-noise ratio, but at somewhat lower redshift and much smaller
angular separation (less than 7.2 arcmin) than our work.  Crotts found
an excess of pairs of $W_0\geq 0.4$ \AA\ lines with $\Delta v < 100$
\kms\ and of $W_0\geq 0.8$ \AA\ lines with $\Delta v < 400$ \kms , but
no clustering for weak ($W_0<0.4$ \AA ) lines.  We find that including
weaker lines to $W_0=0.1$ \AA\ strengthens the clustering
significance.  More data are needed to investigate
such trends and resolve these discrepancies.
We have used our software to re-analyse
published linelists from  smaller, independent datasets
{\cite{Crotts89,Pierre90}}.
These are by far the two largest published
data samples useful for probing correlated QSO \Lya\ absorption
similar to our data.
The two point correlation function similarly reveals
excesses at $50<\Delta v<100$ \kms\
in both of these independent datasets at $2-3\sigma$ significance.

Our key finding is that \Lya\ absorber correlations extend over scales
of 41 arcmin ($\sim 9h^{-1}$ Mpc in the plane of the sky at the $z=3$
frame or 36 comoving $ h^{-1}$ Mpc).
Simulations of the growth
of cosmological
structures{\cite{Hernquist96,Mucket96,Petitjean95,Rauch96b}}
do indicate that filaments of dark matter and gas extend over several
Mpc.  These simulations may
underestimate the true correlation length of the structures traced by
the \Lya\ forest, if it is comparable or larger than the size of the
box in which they are carried out.  Our observations indicate that it
may well be the case, as all the volumes used so far are small compared
to the spatial extent of our survey.

%


%
\vfill
\end{document}